\documentclass{aa}  
\pdfoutput=1 
\usepackage{hyperref}
\usepackage{graphicx}
\usepackage{txfonts}
\usepackage{amsmath}
\usepackage{float}
\usepackage{siunitx}
\usepackage{physics}
\usepackage{pdflscape}
\usepackage{float}
\usepackage{afterpage}
\usepackage{csquotes}
\usepackage{nameref}
\usepackage{pdflscape}
\usepackage{subcaption}
\usepackage[table,xcdraw]{xcolor}

\newcommand{\fdr}{{\tt fdr\,}}

\begin{document} 



\title{On the chaos induced by the Galactic bar on the orbits of nearby halo stars}


\author{{Hanneke C. Woudenberg} 
          \and 
          {Amina Helmi}
          }

   \institute{Kapteyn Astronomical Institute, University of Groningen, Landleven 12, NL-9747 AD Groningen, the Netherlands,}

   \date{Received xxxx; accepted yyyy}

\abstract
  {Many of the Milky Way's accreted substructures have been discovered and studied in the space of energy $E$, and angular momentum components $L_z$ and $L_{\bot}$. In a static axisymmetric system, these quantities are (reasonable approximations of) the integrals of motion of an orbit. However, in a galaxy like the Milky Way with a triaxial, rotating bar, none of these quantities are conserved, and the only known integral is the Jacobi energy $E_J$. This may result in chaotic orbits, especially for inner halo stars. 
   }
   {Here, we investigate the bar's effect on the dynamics of nearby halo stars, and more specifically its impact on their distribution in $(E, L_z, L_{\bot})$ space.}
   {To this end, we have integrated and characterised the orbits of halo stars located within 1 kpc from the Sun. We computed their orbital frequencies and quantified the degree of chaoticity and associated timescales, using the Lyapunov exponent and the frequency diffusion rate.}
  {We find that the bar introduces a large degree of chaoticity on the stars in our sample: more than half are found to be on chaotic orbits, and this fraction is highest for stars on very bound and/or radial orbits. Such stars wander in $(E, L_z, L_{\bot})$ space on timescales shorter than a Hubble time. This introduces some overlap and hence contamination amongst previously identified accreted substructures with these orbital characteristics, although our assessment is that this is relatively limited. The bar also induces a number of resonances in the stellar halo, which are of larger importance for lower inclination, prograde orbits.
    }
   {Because the effect of the Galactic bar on the local halo is important for stars on very bound and/or radial orbits, clustering analyses in these regions should be conducted with care. Replacing the energy by $E_J$ in such analyses could be an improvement.}


\authorrunning{H.C. Woudenberg and A. Helmi}
\titlerunning{Chaos induced by the Galactic bar on nearby halo stars}

\maketitle

\section{Introduction} \label{sec:intro}

A popular method to identify debris from past mergers in our Galaxy is via clustering in the space of energy $E$, and two components of the angular momentum, $L_z$ and $L_\perp$ \citep{Helmi2000}. The argument behind the method first put forward by these authors is that in an integrable, static potential, these quantities are conserved, and therefore a satellite system that is initially compact in position and velocity space, will also be compact in $(E,L_z,L_\perp)$, and remain so even after the system has fully disrupted and phase-mixed. Thus far clustering analyses have been applied relatively successfully to samples of stars near the Sun \citep[see the reviews by][]{Helmi2020,Deason2024}, where our Galaxy has been assumed to be axisymmetric and (very close to) equilibrium. In this case, most stars would have orbits with at least 2 integrals, $E$ and $L_z$,  and while the total angular momentum, and hence $L_{\bot}$, is not an integral it does not vary wildly.  Time dependence due to cosmological growth does not seem to have a very significant effect on these simple assumptions \citep{Gomez2013}, although the debris' complexity is larger, especially for more massive progenitors \citep[e.g.][]{Khopersov2022}. 

There is, however, an additional time dependence and deviation from axisymmetry that is often neglected in such analyses, and that is the presence of the Galactic bar. Whereas its effect on the kinematics of disk stars has been well-established since HIPPARCOS \citep[e.g.][]{Dehnen1999}, its impact on the halo has only recently become clear \citep{Dillamore2023}. Through time-dependence, energy is not conserved, and furthermore the bar introduces resonances, which can induce kinematic substructure \citep{Dillamore2024}. The bar's evident triaxial shape implies that none of the angular moment components are conserved, which may introduce chaoticity \citep{Manos2011}. This will obviously be important for i) stars in the disk and ii) stars on strongly radial orbits that penetrate the innermost regions of the Galaxy.

While it has been widely known that the stellar halo velocity ellipsoid is radially anisotropic, the extent of it became really apparent only with {\em Gaia} data \citep{GaiaDR2}. The halo velocity distribution has a significant population of stars that are on extremely radial orbits, the {\em Gaia}-Sausage \citep{Belokurov2018}, which is associated with the {\em Gaia}-Enceladus merger \citep{Helmi2018}. These stars should be strongly affected by the bar. Stars on the red sequence, \citep[the hot thick disk, or splash population, see][respectively]{Koppelman2018,Belokurov2019} should also be affected by the bar, but differently due to their differing orbital properties.

Here we show that the influence of the bar on the stellar halo near the Sun cannot be neglected. We demonstrate that a large fraction of the halo stars are on chaotic orbits and that therefore the use of $E$, $L_z$ and $L_\perp$ needs some reassessment, particularly for stars on radial and/or very bound orbits. By integrating the orbits of a local sample of halo stars, we study both the degree and effects of chaoticity introduced by a rotating bar on their distribution in $(E, L_z, L_{\bot})$ space.

This paper is structured as follows.  
Sect.~\ref{sec:datamethod} introduces the data and Galactic potential used. In Sect.~\ref{sec:orbits} we describe the orbital characterisation, which includes orbital frequency analysis and the application of chaos indicators. 
In Sect. \ref{sec:results}  we present the results of our analyses. 
We end with a general discussion in Sect.~\ref{sec:disc} and present our conclusions in Sect.~\ref{sec:conclusion}. 

\section{Data and Method} \label{sec:datamethod}

\subsection{Data}
\label{sec:datamethod:generalities}

We use a sample of stars extracted from the \textit{Gaia} DR3 RVS set \citep{GaiaDR3}.  
The distances to the stars were obtained by inverting their parallaxes after correcting them for a zeropoint offset following \cite{Lindegren2021}, and retaining only those for which $\varpi_{\rm corr}/\epsilon_{\varpi} > 5$.  We corrected for extinction using the \cite{Lallament2022} dust extinction maps. We also removed likely globular cluster (GC) members (with probabilities $> 90\%$) from the sample using the membership criteria derived by \cite{Vasiliev2021_GC}. See \citet{Dodd2023} for more details on the quality criteria 
applied. 

To convert to Galactocentric Cartesian and cylindrical coordinates
we also follow \cite{Dodd2023}. Hence, we assume a right-handed coordinated system with the Sun is located in the midplane ($z_{\odot} = 0~\si{\: kpc}$) at $x_{\odot} = -8.2~\si{\: kpc}$ from the Galactic centre \citep{McMillan2017}. To correct for the motion of the Sun, we use $(U, V, W)_{\odot} = (11.1, 12.24, 7.25)$~km s$^{-1}$ \citep{Schonrich2010} and $| \rm \mathbf V_{LSR} | = 232.8$~km s$^{-1}$ \cite{McMillan2017}. The $z$-component of the angular momentum vector, $L_z$, is defined such that it is positive for prograde orbits, meaning that its sign is~flipped. 

To obtain a halo sample we select stars using the Toomre cut $|{\bf V} - {\bf V}_{LSR}| > 210$ ~km s$^{-1}$ and consider only those stars within a distance of 1~kpc from the Sun, and with total velocities smaller than 525~km s$^{-1}$. This velocity cut \citep[well below the estimated local escape velocity, e.g.][]{Koppelman2021v} ensures that all stars are bound in our Galactic potential model (as $E = E_{\rm pot} + E_{\rm kin} \sim -5000$~km$^2$~s$^{-2}$ for this velocity). This leaves us with a sample containing 27885 stars.

\subsection{Milky Way model}

We use \texttt{AGAMA} \citep{Vasiliev2019AGAMA} for orbit integrations and to compute dynamical quantities, and employ the \texttt{AGAMA}  implementation of a barred Galactic potential model as described in \cite{Hunter2024}\footnote{see \href{https://github.com/GalacticDynamics-Oxford/Agama/blob/master/py/example_mw_potential_hunter24.py}{here}}. This is a basis function expansion (BFE) of the analytic model, which allows a higher computational efficiency. This is our fiducial potential. It uses the \cite{Sormani2022} bar model, which is the analytic approximation of the bar model from \cite{Portail2017}. It consists of an X-shaped boxy bulge/bar component, a short bar, and a long bar, and has a total mass of $M_{\rm bar} = 1.83 \cdot 10^{10} M_{\odot}$. Further, it employs the gas disks from \cite{McMillan2017} and an exponential thin and thick stellar disk with a density dip in the centre with parameters that were fit to the \cite{Eilers2019} and \cite{Mroz2019} circular velocity curves. It includes a central black hole, nuclear star cluster, nuclear stellar disk, and a spherical \cite{Einasto1969} dark matter halo with a mass of $M = 1.1 \cdot 10^{12} \: M_{\odot}$. We do not include a potential describing the spiral arms in this work. We set the bar pattern speed equal to $\Omega_b = -37.5$~km s$^{-1}$ kpc$^{-1}$ and assume a bar angle $\phi_b = -25$~degrees, in agreement with recent estimates \citep[see][]{Hunt2025}.

Since the bar is rotating, the Galactic potential is time dependent, and hence a star's total energy is not conserved along its orbit. The Jacobi integral or Jacobi energy $E_J$ may be used instead as it takes this rotation into account. $E_J$ is defined as
\begin{equation}
    \begin{aligned}
     E_J = &\frac{1}{2} \left( v_{x, \rm CR}^2 + v_{y, \rm  CR}^2 + v_{z, \rm CR}^2 \right) + \Phi(x, y, z, t) - \frac{1}{2} \Omega_b^2 R^2,
    \end{aligned}
    \label{eq:EJ}
\end{equation}
where the subscript CR denotes that we are considering quantities in the corotating frame. Positions and velocities in the corotating frame are obtained using 
\begin{equation}
  \begin{aligned}
    \vec{x}_{\rm CR} & = M_R \cdot \vec{x} \\
    \vec{v}_{\rm CR} & = M_R \cdot \vec{v}  - \vec{\Omega_b} \times \vec{x}_{\rm CR}
  \end{aligned}
    \label{eq:coords}
\end{equation}
\noindent  with
\begin{equation}
    M_R = 
      \begin{pmatrix}
    \cos(\phi) & - \sin(\phi)  & 0\\
    \sin(\phi) & \cos(\phi)  & 0\\
    0 & 0 & 1\\
    \end{pmatrix}, 
    \label{eq:rotmat}
\end{equation}
\noindent $\vec{\Omega_b} = (0, 0, \Omega_{b})$, and $\phi = - (\phi_b + \Omega_{b} t)$. Hence, $E_J$ can also be expressed as $E_J = E - \Omega_b L_z$, which shows that its sign does not indicate whether a star is bound or not. 

To ensure that the Jacobi integral is conserved up to at least the order of $10^{-5}$ over the integration time we increase the \texttt{accuracy} parameter in {\tt AGAMA} if necessary. The limiting factor for the conservation of the Jacobi energy in our integrations is the finite smoothness of the BFE interpolator. 

We define the orbital period for each stars, $T_{\rm orb,r}$, as the median time between consecutive apocenters over an integration time of 30 Gyr. The median orbital period for our sample is 120~Myr, with 5\% of the stars having orbital periods smaller than 90 Myr and 5\% having orbital periods larger  than 320 Myr, demonstrating that the stars in our sample have a large range of orbital periods. To make sure that in our analysis each orbit is sampled at a high enough time frequency $\Delta t$ for a long enough period of integration time $T_{\rm int}$, we split up our sample in three subsets:
\begin{itemize}
    \item \texttt{regime 1}: $T_{\rm orb,r} < 0.2$ Gyr, $\Delta t$ = 0.001 Gyr, $T_{\rm int} = $ 15 Gyr, 17741 stars;
    \item \texttt{regime 2}: $ 0.2 < T_{\rm orb,r} < 0.5$ Gyr, $\Delta t$ = 0.003 Gyr, $T_{\rm int} = $ 50 Gyr, 9192 stars;
    \item \texttt{regime 3}: $0.5$ Gyr $ < T_{\rm orb,r} $, $\Delta t$ = 0.005 Gyr, $T_{\rm int} = $ 100 Gyr, 951 stars.
\end{itemize}

\section{Orbital characterisation: frequency analysis, bar resonances, and chaos indicators}
\label{sec:orbits}

\begin{figure}[t!]
\centering
    \includegraphics[width=8.5cm]{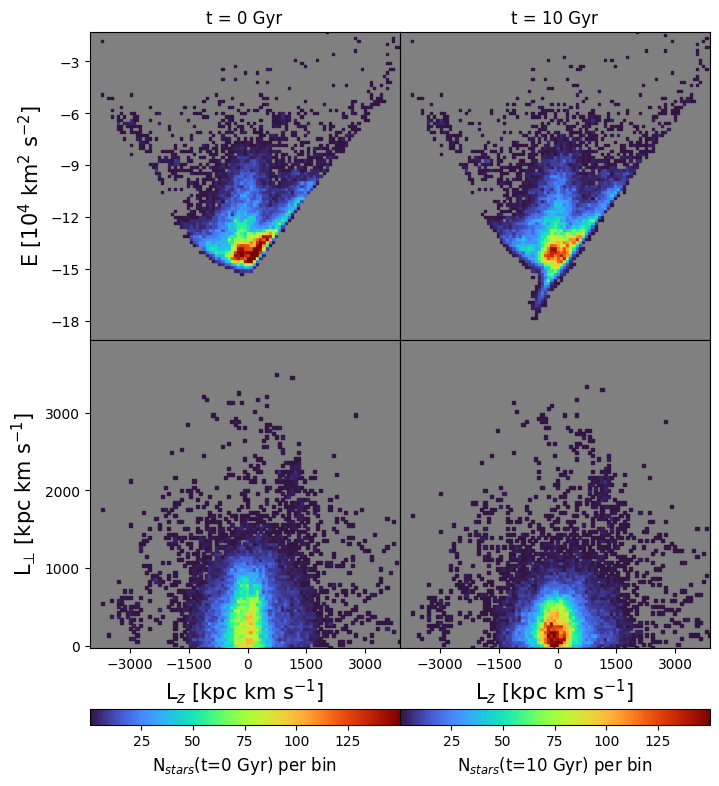}
\caption{\small Distribution of nearby halo stars in $E - L_z$ (top row) and $L_\perp - L_z$ (bottom row) at $t=0$ and after 10 Gyr of evolution in our fiducial Milky Way potential including a rotating bar. Especially highly bound stars with $L_z \sim 0$ (and small $L_\perp$) change their distribution significantly.
\label{fig:IoM_no_IoM}}
\end{figure}

A regular orbit has three integrals of motion (IoM). In the case of (static) axisymmetric systems, the total energy $E$, $L_z$ and $I_3$ are used to denote these integrals, although $I_3$  is often not known explicitly. Therefore, $L_{\bot} = \sqrt{L_x^2 + L_y^2}$ is used as a proxy instead, even though this is not a truly conserved quantity.
In a time-dependent triaxial potential such as the one we use here to represent the Galaxy, none of these quantities are conserved. This is illustrated in Fig. \ref{fig:IoM_no_IoM}, which shows ($E,L_z$) and ($L_z,L_\perp$) computed at $t=0$ and evaluated after 10 Gyr of integration in the fiducial potential, in the top and bottom panels, respectively. Although the distribution of stars globally does not change dramatically\footnote{The general decrease in $L_{\bot}$ is related to the fact that the potential and the observed distribution function are not self-consistent.}, we do see especially large differences for highly bound, low angular momentum orbits. In particular, a fraction of the orbits which initially had $E < -13 000$ km$^2$ s$^{-2}$, have become more bound and mildly retrograde. 

Arguably a regular orbit can be best characterised using its three fundamental orbital frequencies, which are constant in time. 
Generally these cannot be computed analytically but may instead be found by performing a Fourier transform of the complex time series of the orbit's phase space coordinates $\mathbf{x}(t) + i \mathbf{v}(t)$. 
In this work, we consider the orbital frequencies in the corotating frame in Cartesian coordinates (i.e. $\Omega_{x, \rm CR}, \Omega_{y, \rm CR}, \Omega_{z, \rm CR}$) and in a variation of cylindrical coordinates (i.e.  $\Omega_{R,\rm CR}$, $\Omega_{\phi, \rm CR}$, $\Omega_{z, \rm CR}$)\footnote{These sets of orbital frequencies do not necessarily correspond to the orbit's fundamental orbital frequencies, but can be a linear combination of those.}. We use Poincaré symplectic coordinates \citep{Papaphilippou1996, Valluri2012}, in which case the complex time series are of the form $R + i v_{R}$, $\sqrt{2 L_z} \left[ \cos(\phi) + i \sin(\phi) \right]$ and  $z + i v_z$. To determine the orbital frequencies, we used \texttt{SuperFreq} \citep{SuperFreq} with parameters \texttt{min\_freq\_diff = 10**-6}, \texttt{nintvec = 15}, and \texttt{break\_condition = None}. 

\begin{figure}[!h]
\centering
    \includegraphics[width=8cm]{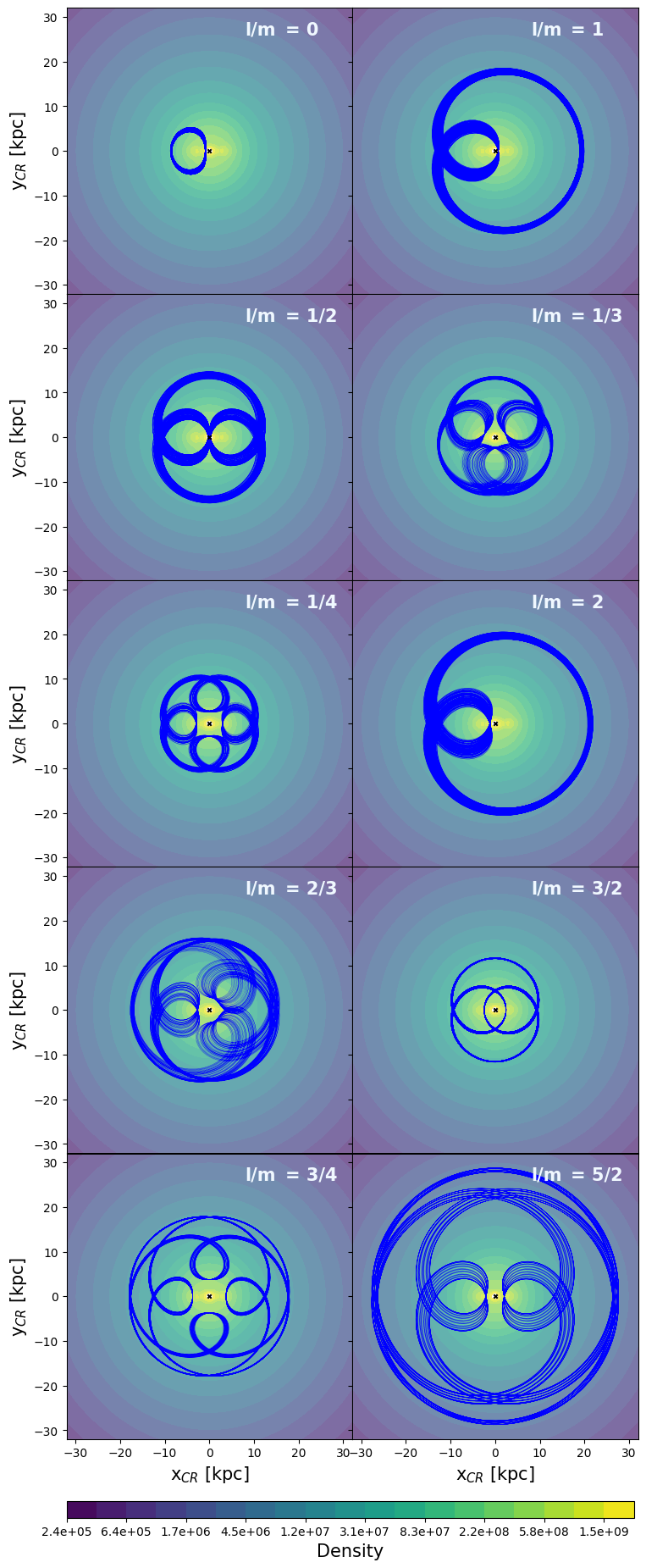}
\caption{\small Examples of orbits of stars on bar resonance, identified as having $\frac{\Omega_{\phi, \rm CR}}{\Omega_{R, \rm CR}} = l/m \pm 0.005$ and $\lambda = 0$ and \fdr < -1.9. The orbits are shown in the corotating frame, and the resonance they are associated to is indicated in the top right corner. Isodensity contours of the fiducial potential in the midplane ($z = 0$) are shown in the background, the black cross indicates the Galactic Center.}
\label{fig:orbits}
\end{figure}

If two or three of the orbit's fundamental frequencies are a small integer ratio of one another, the corresponding orbit is said to be resonant. In a rotating barred potential, an orbit can also be in resonance with the bar, and this occurs when
\begin{equation}
    m (\Omega_{\phi} - \Omega_b) + l \Omega_R = 0
    \label{eq:barres}
\end{equation}
in the inertial frame, or 
\begin{equation}
    m (\Omega_{\phi, \rm CR}) + l \Omega_{R, \rm CR} = 0
    \label{eq:barres_CR}
\end{equation}
in the corotating frame, with $l$ and $m$ being an integer and where $\Omega_b$ is the bar's pattern speed. $l/m = 0$ corresponds to the corotation resonance (an orbit has the same azimuthal orbital frequency as the bar), $l/m = -1/2$ corresponds to the inner Lindblad resonance (ILR), and $l/m = 1/2$ corresponds to the outer Lindblad resonance (OLR).

Fig. \ref{fig:orbits} shows examples of orbits in resonance with the bar. These have been selected using $\frac{\Omega_{\phi, \rm CR}}{\Omega_{R, \rm CR}} = l/m \pm 0.005$ for a range of integers $l$ and $m$. 
Their present-day $L_{\bot}$ and $L_z$ values range from 0 to 900~kpc~km~s$^{-1}$, and from -1000 to 1500~kpc~km~s$^{-1}$, respectively. 
This figure shows that even orbits with large apocenters (e.g. 20 or 30 kpc) can be in resonance with the bar.

\begin{figure}
\centering
    \includegraphics[width=\hsize]{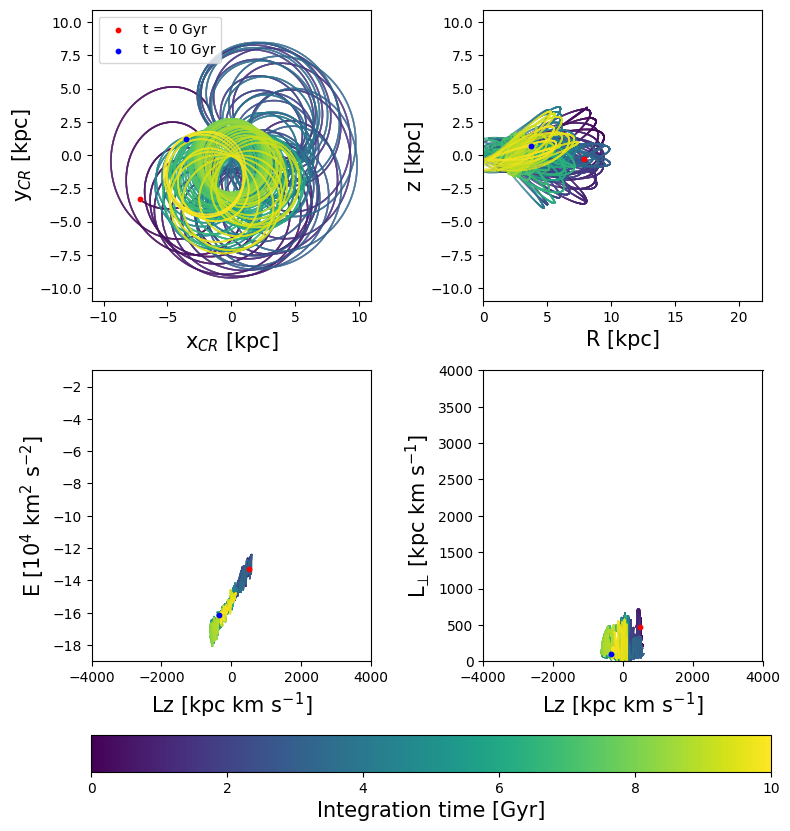}
\caption{\small Example of a star on a chaotic orbit (selected from \texttt{regime1}). 
The red and blue markers show the position of the star at the present day and after 10 Gyr of integration time, respectively. The top panels show projections of the orbit in the corotating frame, ($x_{\rm CR}$,  $y_{\rm CR}$), and in $(R, z)$, respectively. The bottom panels illustrate the trajectory of this star in $(E, L_z$, $L_{\bot})$ space over time. This star ends up librating around the bar's corotation resonance after 10 Gyr, and moves around significantly in $(E, L_z, L_{\bot})$ space as it changes its orbit. The orbit has $\lambda = 2.1$~Gyr$^{-1}$ and $t_{\lambda}$ = 1.3 Gyr, and \fdr = $-0.2$ (see Sect.~\ref{sec:fdr} and \ref{sec:Lyap} for details on how these quantities are computed).}
\label{fig:orbit_example}
\end{figure}

Chaotic orbits have fewer than three IoM and their orbital frequencies thus change over time. An example of a chaotic orbit is shown in Fig. \ref{fig:orbit_example}. After 10~Gyr of integration time, this orbit seems to be librating around the corotation resonance, and it changed its apocenter but also $E, L_z$, and $L_{\bot}$ by a substantial amount. Clearly, in the case of this orbit, the bar is able to cause chaotic behaviour that manifests itself after only a few Gyr. 

\subsection{Chaos indicators: frequency diffusion rate}
\label{sec:fdr}

To quantify the degree and effect of chaoticity in our halo sample, we make use of the Lyapunov exponent $\lambda$, and we calculate the frequency diffusion rate, which we denote as \fdr. These are different, but complementary measures of chaoticity. To obtain a chaoticity timescale we use the Lyapunov exponent. This timescale gives an indication of how quickly an orbit changes due to chaos. 

The \fdr is based on the fact that the orbital frequencies change over time for a chaotic orbit. To compute it, we take the difference in orbital frequency between two consecutive periods of equal integration time, $0 < t_1 < T_{\rm int}$ and $ T_{\rm int} < t_2 < 2 \: T_{\rm int}$, and determine 
\begin{equation}
    \Delta f_i  =  \left|   \frac{\Omega_i(t_1) - \Omega_i(t_2)}{\Omega_i(t_1)}  \right| 
    \label{eq:freqdiffrate}
\end{equation}
for each of the three frequencies $\Omega_i$. We take the geometric average of the three $\Delta f_i$ and then its logarithm to obtain 
\begin{equation}
    \text{\tt fdr} = \log_{10} \Delta f, \;\; {\rm where} \: \Delta f = \left[ \prod_{i} \Delta f_i  \right]^{\frac{1}{3} }.
    \label{eq:fdr_geom}
\end{equation}  
The larger the value of the \fdr, the more chaotic the orbit is. However, in general there is no clear threshold value of the \fdr that separates chaotic from regular orbits, and instead, its spectrum is continuous \citep{Valluri2010}. Nonetheless, we find that the \fdr correlates well with the Lyapunov exponent $\lambda$ \citep[see also][and Sect.~\ref{sec:results:degreechaos}]{Vasiliev2013}. 
An important caveat is that for resonant orbits the frequencies are not independent, and consequently the determination of the \fdr becomes less reliable in such cases.

\subsection{Chaos indicators: Lyapunov exponent}
\label{sec:Lyap}

The Lyapunov exponent measures the evolution in time of the infinitesimal 6D deviation vector $\vec{w}(t)$ from a given orbit. 
The divergence of nearby orbits in configuration space grows linearly in time for a regular orbit \citep{Helmi1999,Gomez2007,Vogelsberger2008}, and exponentially in the chaotic case \citep{Maffione2015, PriceWhelan2016}, and therefore we expect a similar behaviour for the deviation vector. In this work, we use the set of 6 deviation vectors computed by {\tt AGAMA} to derive the Lyapunov exponent and the onset time\footnote{A similar method to calculate $\lambda$ and t$_{\rm onset}$, building on the approach presented in this work, will become available in {\tt AGAMA} (Vasiliev, private communication).} of the exponential growth regime, $ t_{\rm onset}$. To this end, we take the norm of each deviation vector per timestep, and select the vector with the largest norm over time, $\max \left( \Vert{\vec{w_1}_{\rm}}(t)\Vert, ..., \Vert{\vec{w_6}_{\rm}}(t)\Vert \right)$. We then 
perform a non-linear least squares fit using  \texttt{scipy.curve\_fit} of the following function, 

\begin{equation}
\ln\left(\frac{\Vert{\vec{w}_{\rm max}}(t)\Vert}{t} \right) = \left\{
    \begin{array}{ll}
        C & {\rm for} \: \:  t < t_{\rm onset} \\
        C + \lambda (t - t_{\rm onset}) & {\rm for} \: \: t > t_{\rm onset},
    \end{array}
\right.
\label{eq:HShiLloL}
\end{equation}
\noindent where $C$ is a constant. If there is no exponential growth, $\lambda = 0$ and $t_{\rm onset}$ is set to be equal to the integration time, while if there is exponential growth, ${\lambda}$ is returned in units of Gyr$^{-1}$ and $t_{\rm onset}$ will be smaller than the integration time and zero if chaotic behaviour starts immediately. To make sure chaoticity is robustly detected, we set $\lambda = 0$ if $T_{\rm int} - t_{\rm onset} < 5 T_{\rm orb}$, implying that the time of exponential growth is detected over a period shorter than 5 times the orbital timescale, or if $ (T_{\rm int} - t_{\rm onset}) \lambda < 0.1 \ln(T_{\rm int} )$, that is if the increase in the norm of the deviation vector due to chaoticity is smaller than 10\% of the growth expected from the divergence for a regular orbit over the integration time.  We have checked that the method implemented via Eq.~\ref{eq:HShiLloL} gives similar values for $\lambda$ as when computed using an independent method \citep{Voglis2002}, and that the derived magnitude of $t_{\rm onset}$ roughly agrees with when the orbit starts to depict exponential divergence. To make sure $t_{\rm onset}$ is reliably determined within a Hubble time, we integrated stars in the subset defined as \texttt{regime1} for 25 Gyr when computing their deviation vectors.

\begin{figure*}[t]
\centering
\includegraphics[width=17cm]{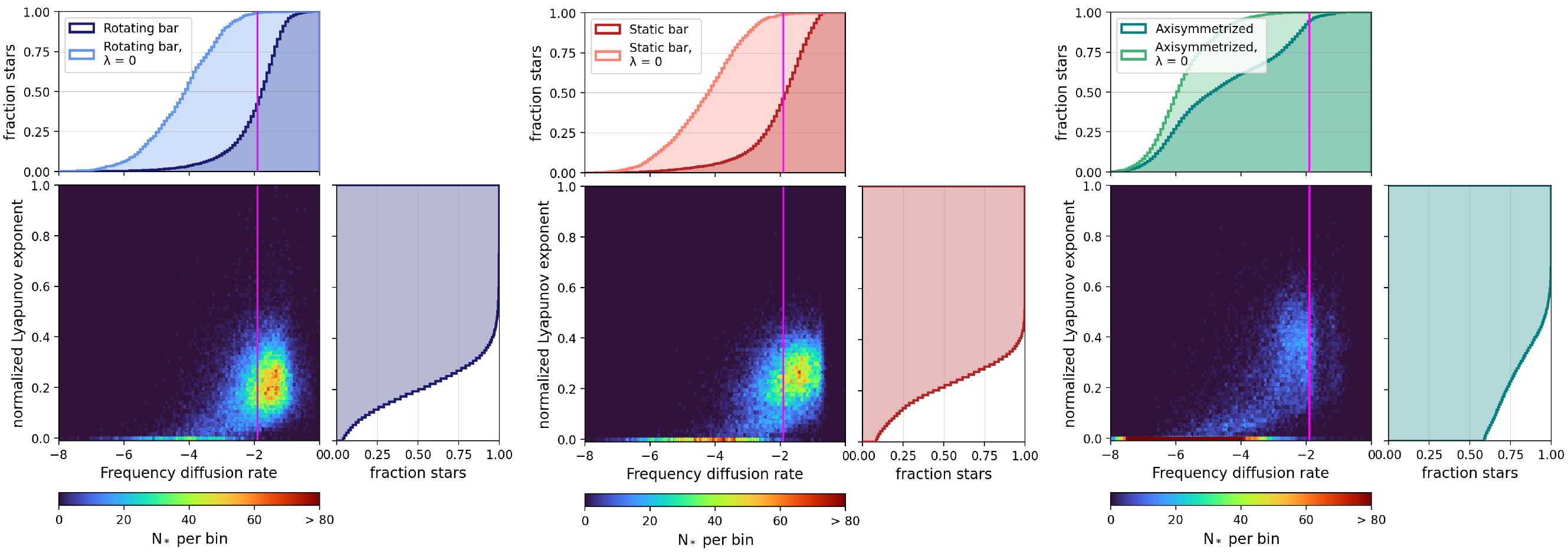}
\caption{\small Distribution of the values of the two chaoticity indicators used in this work for our local halo sample: the Lyapunov exponent $\lambda$, normalized by $T_{\rm orb, r}$, and the \fdr. The different panels correspond to three different versions of the potential described in Section \ref{sec:datamethod:generalities}: with a rotating bar (left), with a static bar (middle), and an axisymmetrised bar (right). The side panels show the cumulative distributions of normalized $\lambda$ for each of the potentials, while the top panels correspond to those of \fdr. The \fdr distribution for the orbits that have $\lambda = 0$ is shown separately. The vertical magenta lines correspond to \fdr = -1.9, the 99th percentile of the orbits with $\lambda = 0$ in the fiducial potential. }
\label{fig:distribution_lyap_fdr}
\end{figure*}

We define a timescale of chaoticity $t_\lambda$, using both the Lyapunov exponent and the chaos onset time as
\begin{equation}
    t_{\lambda} =  \frac{1}{\lambda} + t_{\rm onset}.
    \label{eq:lyapunovtime}
\end{equation}
We generally find that $t_\lambda$ gives a lower limit to the timescale over which chaoticity is manifested macroscopically (i.e. a significant change in the orbit that is apparent through visual inspection) and that typically $5/\lambda + t_{\rm onset}$ better reflects the time over which chaoticity becomes visually discernible.

\section{Results} \label{sec:results}

We present here the analysis of the orbits of our local sample of halo stars. We first focus on the distribution of their Lyapunov exponents and \fdr, and then identify where in the present-day's $(E, L_z, L_{\bot})$  space and for what types of orbits the effect of chaoticity is largest. We also locate the bar resonances and determine the fraction of stars trapped on or affected by these. Finally we discuss how substructures identified in previous work in the present-day's space of $(E, L_z, L_{\bot})$ may be affected by the bar. 

\subsection{Degree of chaoticity for halo stars near the Sun}
\label{sec:results:degreechaos}

Figure \ref{fig:distribution_lyap_fdr} shows the distribution of the values of the two chaoticity indicators introduced in the previous section for our local halo sample: the Lyapunov exponents and the \fdr. We show the Lyapunov exponents normalized by $T_{\rm orb, r}$ to allow for a more direct comparison between orbits with different orbital timescales. The indicators have been computed for three variations of the potential described in Section \ref{sec:datamethod:generalities}: the fiducial potential itself, which is triaxial and time-dependent (left panels), a model with a static bar (middle panels), and an axisymmetrised model of the fiducial potential, meaning that in the basis function expansion only the $m=0$ term is retained, making it also time-independent (right panels).

As expected, the largest number of stars on regular orbits are found for the axisymmetric potential, where 56\% of the stars have $\lambda = 0$. A much smaller fraction of the stars have $\lambda = 0$ for the rotating and static bar potential, 4\% and 8\%, respectively. The distributions for these two triaxial cases are relatively similar, with the \fdr increasing sharply for \fdr$> -2$ (generally corresponding to normalized $\lambda > 0.1$). The similarity of these distributions indicates that the chaoticity is driven mostly by the triaxiality present in the potential and to a smaller degree by the resonances with the rotating bar. 

To obtain a robust estimate of the amount of stars on chaotic orbits
we prefer to use both the \fdr and Lyapunov exponent. To identify such
orbits we require $\lambda > 0$ and $\text{\tt fdr} > -1.9$, since
this value corresponds to the 99th percentile of the distribution of
\fdr for orbits that have $\lambda = 0$ in the fiducial
potential\footnote{We checked that the few orbits above this limit are
  either regular or close to an orbital resonance, in which case the
  \fdr and Lyapunov exponent determinations are less reliable.}, as can
be inferred from the top sub-panels of
Fig.~\ref{fig:distribution_lyap_fdr}. With these criteria we find that
60\% of the stars in our sample are on chaotic orbits in the rotating
bar potential, and a comparable percentage, namely 56\%, in the static
bar potential. Blind application of this criterion to the axisymmetrized potential would
yield that only 7\% of the stars are on chaotic orbits. The right panel of 
Fig.~\ref{fig:distribution_lyap_fdr} shows this would be an incorrect assessment, because for orbits with $\lambda = 0$ the 99th percentile takes a lower \fdr value, namely $-3$.  In fact, by inspecting the various indicators and the chaos onset times 
we find that the region with $\lambda < 0.15$ and \fdr < $-1.9$ hosts sticky chaotic orbits with large chaos onset times, which comprise 13\% of the stars. Stars on fully chaotic orbits would be those with 
($\lambda > 0.15$ and $-3 <$ \fdr $<-1.9$) or ($\lambda > 0$ and \fdr > $-1.9$), and comprise
26\% of the sample.

\begin{figure*}[t!]
\centering
    \includegraphics[width=0.9215\hsize]{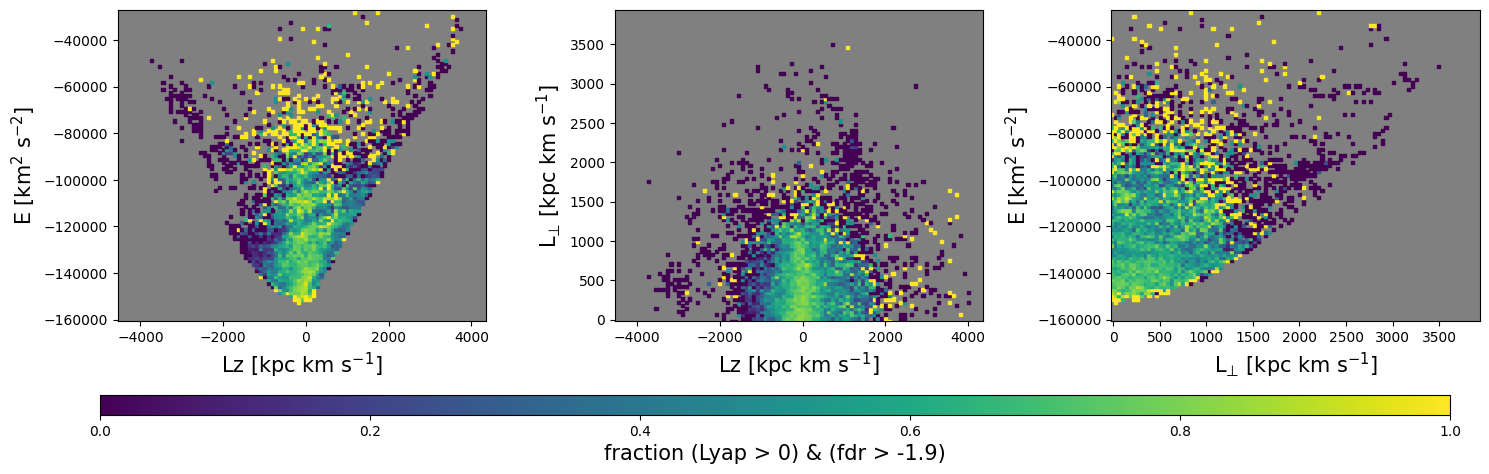}  
\caption{\small Local halo stars in $(E, L_z, L_{\bot})$ space  computed at $t=0$ in the fiducial potential, binned into 100 bins along each axis, where the colour coding of each bin shows the percentage of stars with a Lyapunov exponent higher than 0 and a \fdr $ > -1.9$, determined after 25 Gyr of integration.}
\label{fig:Lyap_IoM}
\end{figure*}

\begin{figure*}
\centering
\begin{subfigure}{0.97\textwidth}
    \centering
    \includegraphics[width=.95\linewidth]{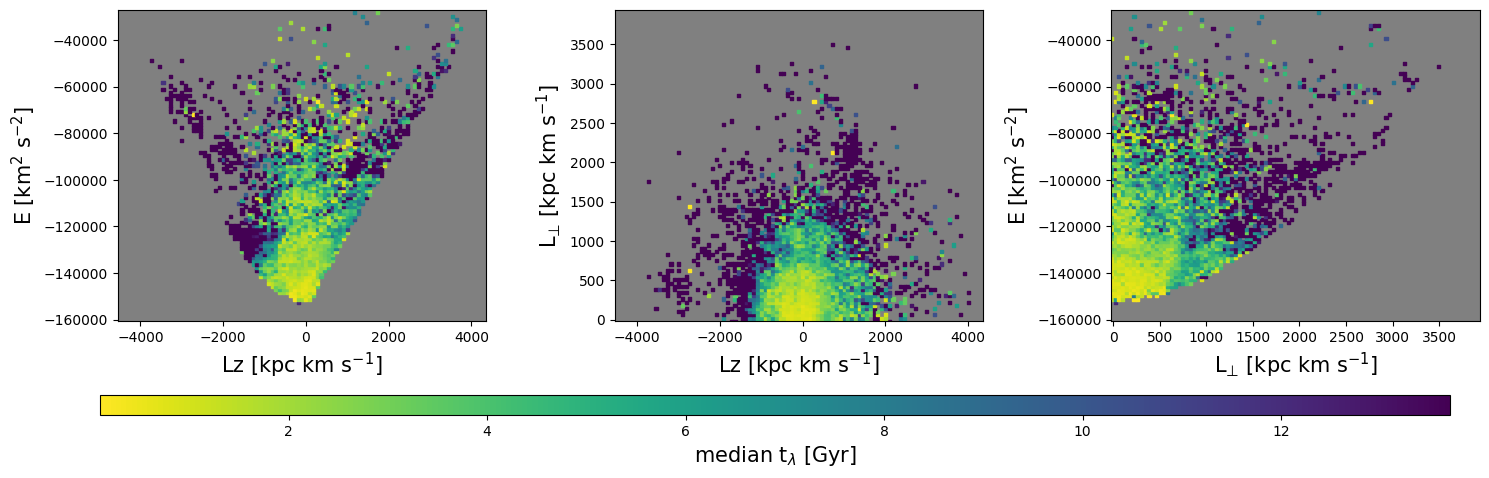}  
\end{subfigure}
\begin{subfigure}{0.97\textwidth}
    \centering
    \includegraphics[width=0.95\linewidth]{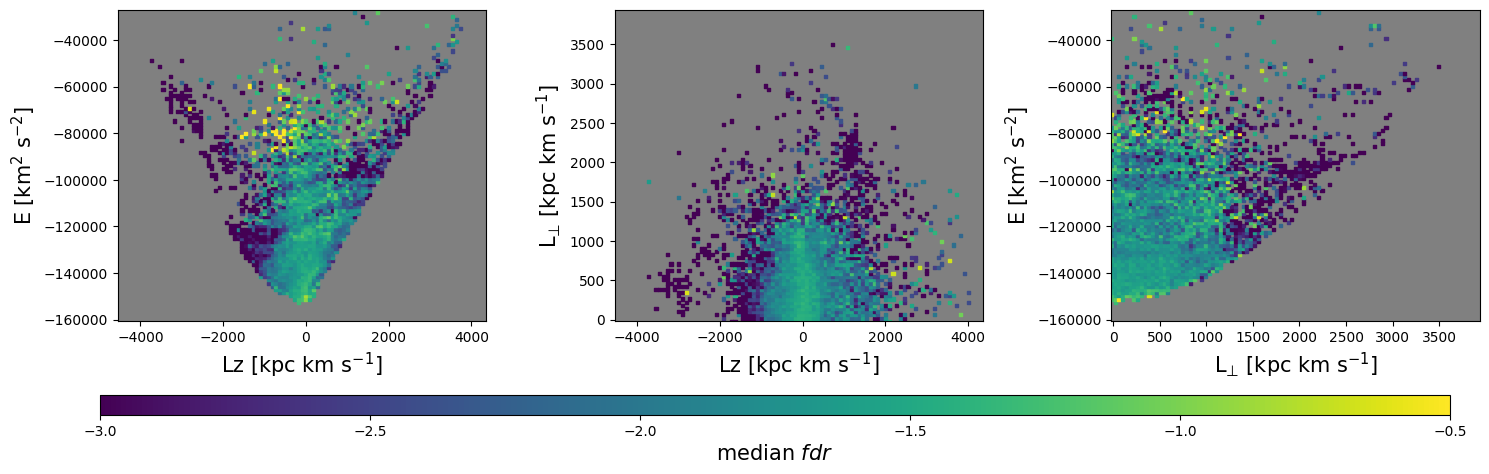}
\end{subfigure}
\caption{\small As Fig.~\ref{fig:Lyap_IoM}, now showing in the {\bf top panel:} the weighted median Lyapunov timescale (Eq. \ref{eq:lyapunovtime}), and in the {\bf bottom panel:} the median {\tt fdr} (Eq. \ref{eq:fdr_geom}) determined over an integration time of 25 Gyr for all stars in the sample. Note the good correspondance between both panels, and with Fig.~\ref{fig:Lyap_IoM}.}
\label{fig:Lyap_fdr_timescale_IoM}
\end{figure*}

\subsection{Chaoticity in the present-day space of $E$, $L_z$ and $L_\perp$}
\label{sec:results:timescale}

We investigate here which region of present-day $(E, L_z, L_{\bot})$  space and hence what types of orbits for the stars in our sample are most affected by the presence of a (rotating) bar. Fig. \ref{fig:Lyap_IoM} shows the percentage of stars that have Lyapunov exponents higher than 0 and \fdr$~>~-1.9$ per bin in $(E, L_z, L_{\bot})$ space. Clearly, stars on radial orbits, having $L_z \sim 0$ kpc km s$^{-1}$,  and/or strongly bound orbits are for the large majority chaotic. About 70\% of the stars in our halo sample with $|L_z| < 500$ kpc km s$^{-1}$ have chaotic orbits, and this percentage increases by a few percent for stars on more bound orbits ($E < -140 000$ km$^2$ s$^{-2}$). Instead, stars on the retrograde side of the halo, having $L_z \leq -1000$ kpc km s$^{-1}$, or stars with high inclination orbits, having $L_{\bot} \geq 1500$ kpc km s$^{-1}$, are barely affected by the bar. 

We explore now how the Lyapunov timescale $t_{\lambda}$ (Eq. \ref{eq:lyapunovtime}) varies as a function of the present day location of stars in $(E, L_z, L_{\bot})$ space. The top row of Fig.~\ref{fig:Lyap_fdr_timescale_IoM} shows the median $t_{\lambda}$  per bin in $(E, L_z, L_{\bot})$ space, while the bottom row shows the median \fdr. The two chaos indicators show good agreement in terms of structure in $(E, L_z, L_{\bot})$ space, indicating that chaoticity is detected robustly, as the two measures are independent. The top panels show that   there are regions where chaos acts on a timescale (much) shorter than the Hubble time, especially for orbits that are strongly bound and/or radial, which can significantly change even within a couple of Gyr or less. 

\begin{figure*}[!t]
\centering
  \includegraphics[width=\textwidth]{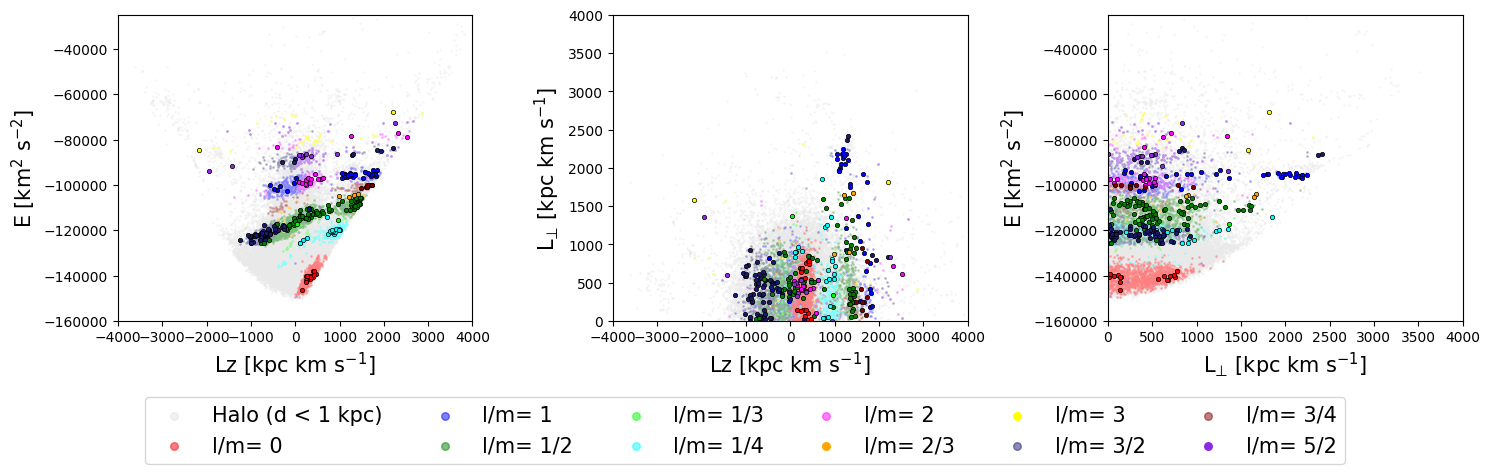}  
\caption{\small Position of local halo stars close to bar resonances in $(E, L_z, L_{\bot})$  space at the present day, selected to have  $\frac{\Omega_{\phi, \rm CR}}{\Omega_{R, \rm CR}} = l/m \pm 0.005$ (see Fig. \ref{fig:orbits} for examples of each resonant orbit). The stars encircled in black are on resonant bar orbits, i.e. with $\lambda$ = 0 and \fdr < -1.9.}
\label{fig:bar_resonances}
\end{figure*}

These short timescales are accompanied by a large change in $E, L_z$ and $L_{\bot}$, as was also illustrated in Figs. \ref{fig:IoM_no_IoM} and \ref{fig:orbit_example}. Fig. \ref{fig:orbit_example} also illustrates the usual path that an initially relatively strongly bound, very radial orbit takes over time in $(E, L_z, L_{\bot})$ space: namely, along a line of positive slope in $(E, L_z)$, and along a horizontal line with a large amount of variation in $(L_z, L_{\bot})$, generally reaching $L_{\bot} = 0$ kpc km s$^{-1}$ \citep[see also][]{Dillamore2024}. Due to this, within the timespan of a few Gyr, a star's orbit can change from being prograde to retrograde (change in $L_z$), from being confined to the disk to reaching values of $z_{\rm max}$ of a few kpc (as a result of a change in $L_{\bot}$) and from passing through the local volume to not being able to reach that volume at all because the apocenter has shrunk (change in $E$). 
As Figs.~\ref{fig:Lyap_IoM} and \ref{fig:Lyap_fdr_timescale_IoM} show, this mixing in $(E, L_z, L_{\bot})$ space is particularly strong for very bound, relatively radial orbits. 
This has implications for the reliability of the identification of substructure associated to merger debris, and it also implies that substructures located today in such regions of this space could suffer significant contamination. We return to this point in Sect.~\ref{sec:effect_substr}.

\subsection{The presence of resonances}
\label{sec:results:resonances}

The stripy diagonal structures seen in Figs.~\ref{fig:Lyap_IoM} and
\ref{fig:Lyap_fdr_timescale_IoM} can be linked to the bar resonances
that are present in this region. This is illustrated in
Fig.~\ref{fig:bar_resonances}, which shows stars that have been
selected to have
$\frac{\Omega_{\phi, \rm CR}}{\Omega_{R, \rm CR}} = l/m \pm 0.005$ for
a range of $l$ and $m$. Approximately 13\%
of the stars in our sample are sufficiently close to a resonance, meaning that they satisfy this
condition. Note that this criterion selects stars that are very
close but not necessarily exactly on the resonance. Since resonant
orbits should be regular, we impose a second criterion to select these, namely, we also
require that $\lambda = 0$ and \fdr < -1.9. This additional
criterion separates these from orbits trapped by
resonances temporarily while being intrinsically chaotic. 

Fig.~\ref{fig:bar_resonances} shows the position of the stars in today's $(E, L_z, L_{\bot})$  space selected according to their frequencies as described in the previous paragraph, with each bar resonance shown in a different colour. Stars on bar resonances are shown as symbols with black edgecolours, and can be seen to form a more compact set in the space. Clearly, the stripes apparent in Fig. \ref{fig:Lyap_IoM} and \ref{fig:Lyap_fdr_timescale_IoM} (which have larger median $t_\lambda$ and smaller \fdr values) correspond to regions hosting resonant bar orbits (which are stable). These results qualitatively agree with the findings by \citet{Dillamore2024}, who used a different gravitational potential hosting a bar. We also find that if the bar is non-rotating such stripes are not present. 

The region marking the transition from a specific bar resonance to another orbit is known as separatrix, and orbits located in this region are prone to chaotic behaviour, as is the case for the chaotic orbit shown in Fig. \ref{fig:orbit_example}. In fact, we find that the vast majority of stars selected according to the orbital frequency criterion to be close to a bar resonance have positive Lyapunov exponents, with roughly 50\% having $\lambda > 1$ Gyr$^{-1}$. 
Such stars essentially move along the inclined lines in $(E, L_z)$ space delineated by the bar resonances seen in Fig. \ref{fig:bar_resonances}. 

The above analysis reveals that the highly chaotic regions essentially have two sources. The most important cause for chaotic behaviour is the presence of a triaxial bar in the center of the Galaxy, which importantly affects stars that pass close to it (those on radial and/or bound orbits), even if this bar is static (see Section \ref{sec:results:degreechaos}). The second source of chaoticity are the separatrices of the bar resonances as described above, which affect to a larger degree the stars part of the hot thick-disk, as they are on prograde orbits of low inclination.

\subsection{Impact on substructures}
\label{sec:effect_substr}

\begin{figure*}[!t]
\centering
\begin{subfigure}{1\textwidth}
    \centering
    \includegraphics[width=1\linewidth]{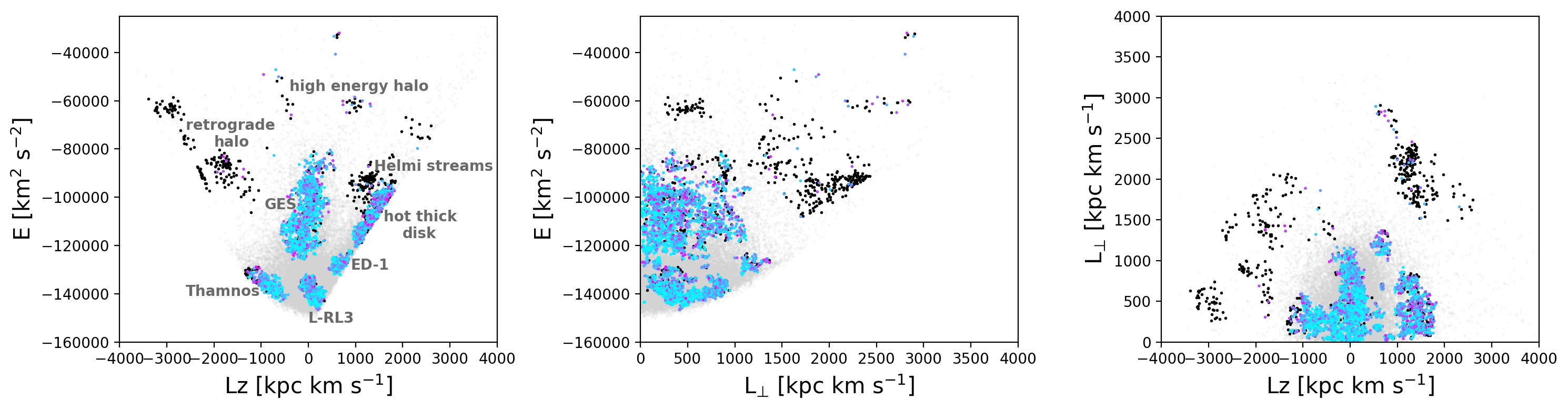}  
\end{subfigure}
\begin{subfigure}{1\textwidth}
    \centering
    \includegraphics[width=1\linewidth]{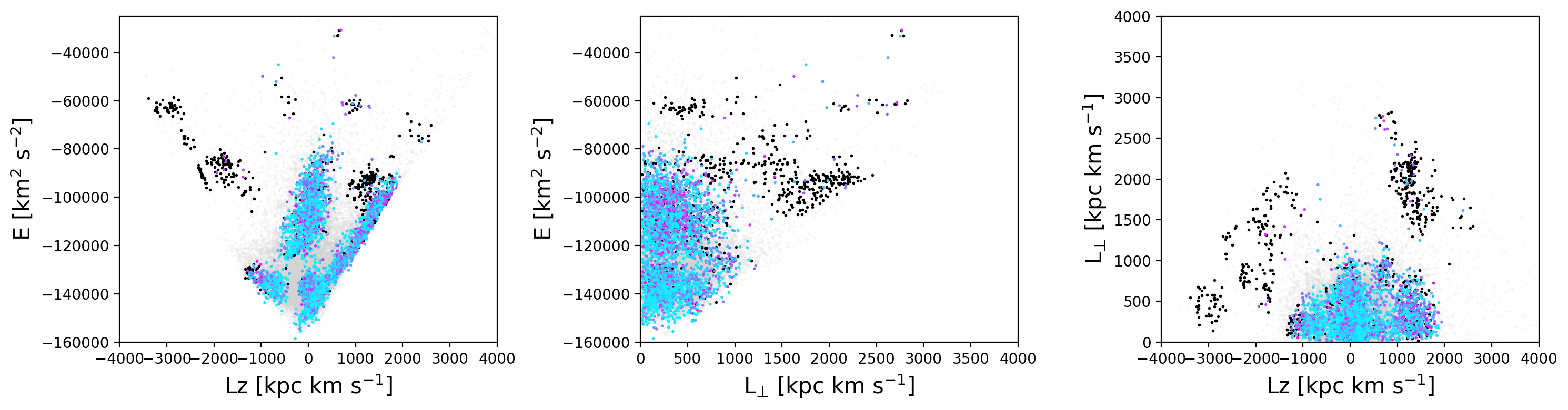}
\end{subfigure}
\begin{subfigure}{1\textwidth}
    \centering
    \includegraphics[width=0.66\linewidth]{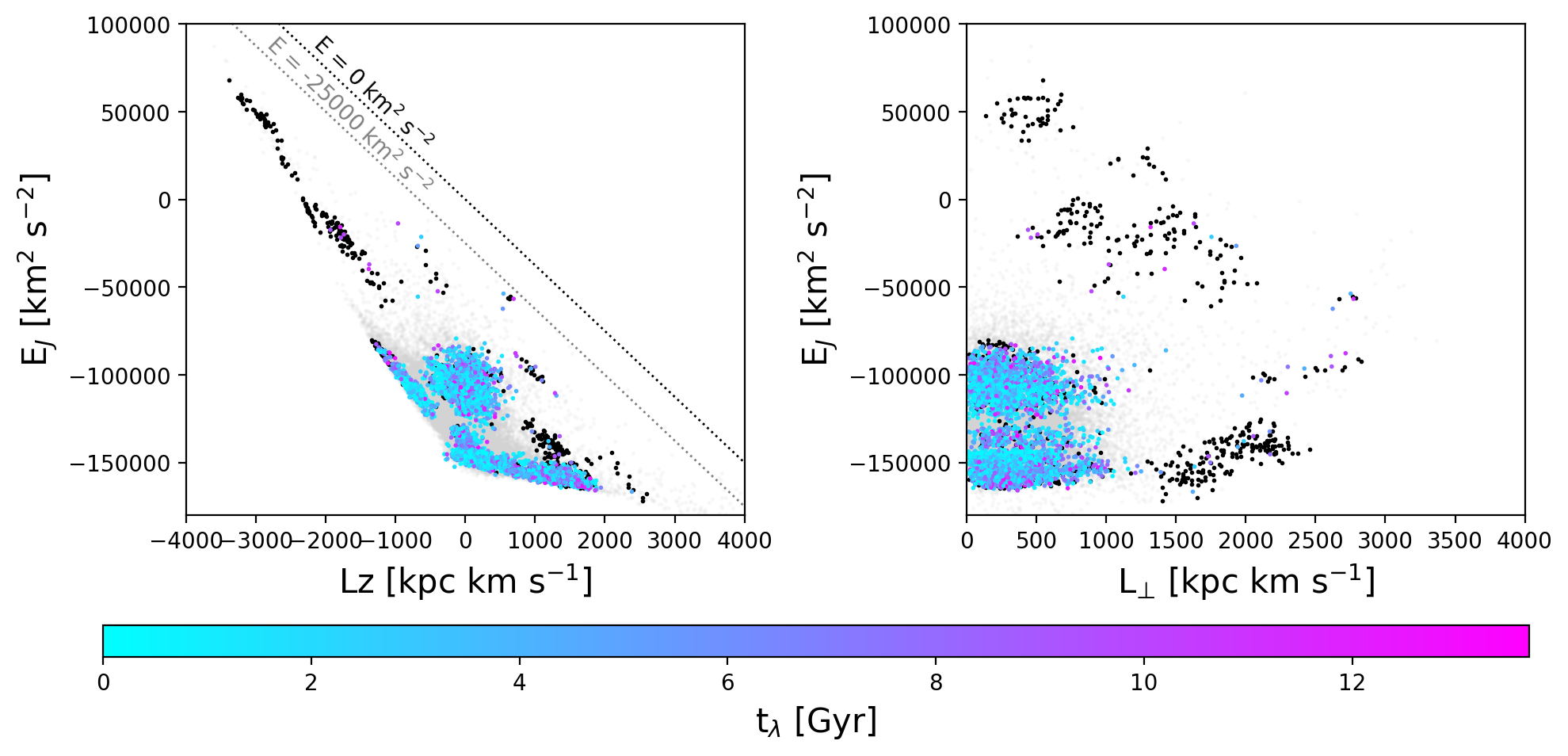}
\end{subfigure}
\caption{\small \textbf{Top panel:} Stars part of the substructures identified by \cite{Dodd2023} in $(E, L_z, L_{\bot})$ space at the present day, colour-coded by $t_{\lambda}$. Stars with $t_{\lambda}>13.7$~Gyr are shown in black. The light grey stars in the background correspond to the whole local halo sample. A number of substructures and regions of $(E, L_z, L_{\bot})$ space are labelled in the leftmost panel. \textbf{Middle panel:} Same as the top panel, but after 10~Gyr of integration time in the fiducial potential, and only showing stars whose final three apocenters were on average larger than 7~kpc. \textbf{Bottom panel:} Also after 10 Gyr of integration time, but showing $(E_J, L_z)$ and $(E_J, L_{\bot})$ instead of $E$. We have indicated the line $E_J = E - \Omega_b L_z$ for $E = -25000$~km$^2$~s$^{-2}$ (corresponding to the upper limit in energy of the top and middle panel) and $E = 0$~km$^2$~s$^{-2}$  (corresponding to unbound stars in our potential).}
\label{fig:substructures}
\end{figure*}

In Fig. \ref{fig:substructures} we show the stars associated to the
substructures identified by \cite{Dodd2023} colour-coded by their
$t_{\lambda}$. If a star's $t_{\lambda}$ is larger than a Hubble time,
it is shown in black. The top panel of this figure shows the
present-day distribution in $(E, L_z, L_{\bot})$ space, while the
middle panel shows the distribution after 10 Gyr of integration time
in the fiducial rotating bar potential. There, we only show the stars
whose final three apocenters were on average larger than 7~kpc,
meaning they should still be able to reach or cross a volume of
$\sim 1$~kpc radius centred on the Solar circle. This requirement removes a bit more than 4\% of the stars, which are on radial and bound orbits close to the corotation resonance.

At first glance, the main difference between the top and middle panels
is that the distributions of substructures with $t_{\lambda}$ smaller
than a Hubble time are smeared out. More specifically, the hot thick
disk, ED-1 and L-RL3 are now connected in $(E, L_z)$ space. We note, however, that the nature of the latter two substructures has never
been clear, for instance their chemistry reveals a mix of populations \citep{RuizLara2022, Dodd2023}, so perhaps this is not so unexpected in hindsight. The most bound stars of {\it Gaia}-Enceladus form a tail
towards lower energy and more negative $L_z$ after 10 Gyr, meaning that they moved along the resonance
lines described in the previous section, and get closer to the region
occupied by Thamnos. The stars associated to the latter substructure cover a range
of regimes in $t_\lambda$, from short ($\sim 3$~Gyr) to greater than a Hubble
time, which is the case for orbits with $L_z < -1000$~kpc km s$^{-1}$. 

Substructures that are on less bound orbits (high-energy halo), 
or have $L_z < -1000$~kpc km s$^{-1}$ (retrograde halo), or with significant $L_\perp$ ($\gtrsim 1500$~kpc km s$^{-1}$) and hence reaching high above the mid-plane,  
are generally on regular orbits. 
These stable regions include structures like the Helmi Streams, Sequoia, RLR-64/Antheus, ED-2/3/4/5/6, and Typhon.

Given the results presented here, it is clear that a rotating bar
affects the distribution, coherence, and purity of substructures in
significant parts of $(E, L_z,
L_{\bot})$ space. Therefore for any clustering analysis, it would be preferable to 
consider quantities that remain conserved with time. Hence, instead of
using the total energy, it might be interesting to use the Jacobi 
energy $E_J$ (Eq. \ref{eq:EJ}).  
The bottom panel of
Fig. \ref{fig:substructures} shows a projection of the
substructures in $(L_z, E_J)$, $(L_{\bot},
E_J)$ after 10 Gyr of integration time in the fiducial 
potential with the rotating bar. We see that substructures that are not strongly affected by the
bar separate from those with short Lyapunov
timescales, especially in $(L_{\bot},
E_J)$ space. There, almost all stars part of substructures with high
$L_{\bot}$ and/or high
$E_J$ have Lyapunov timescales longer than a Hubble time. We note that, because of
the somewhat squashed appearance of the ($E_J, L_z$) subspace, it might be
advantageous to do some scaling before applying a clustering algorithm in $(E_J, L_z,
L_{\bot})$ space.

\section{Discussion} 
\label{sec:disc}

\begin{figure*}
    \centering
    \includegraphics[width=\textwidth]{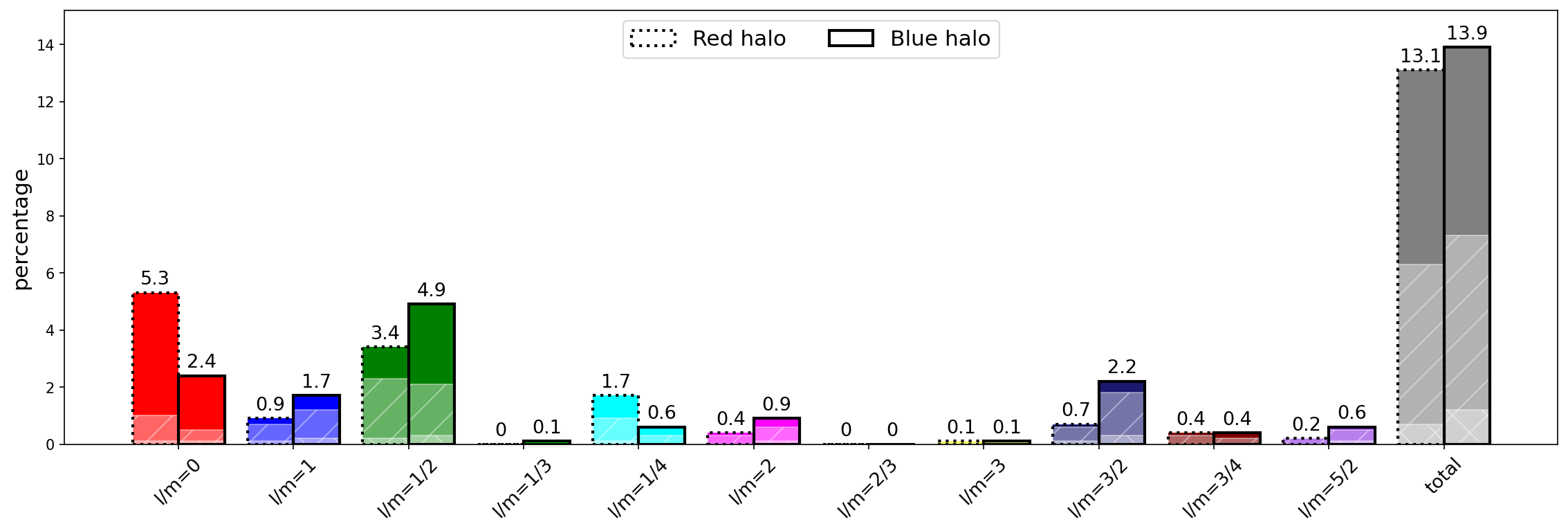}
\caption{\small Percentage of local halo stars close to  different bar resonances for the red halo (dashed outlines) and red halo (solid outlines). The percentages on top of each bar corresponds to all selected stars, the white hatched bar indicates the percentage of stars with a Lyapunov exponent < 1 Gyr$^{-1}$ (meaning the non-hatched part of the bar corresponds to stars with Lyapunov exponent > 1 Gyr$^{-1}$ ), and the doubly hatched bar indicates the percentage of stars with a Lyapunov exponent equal to zero.}
\label{fig:bar_resonances_hist}
\end{figure*}

\subsection{Robustness under different (bar) potential models}

The results derived here hold for the fiducial potential, with it its
specific bar model, but many bar parameters such as its length, mass,
pattern speed and angle are still somewhat uncertain (see
\cite{Hunt2025} for an overview). To quantify how much changing these
parameters affects our results, we have explored two different bar
models (see Appendix ~\ref{sec:appendix:barmodels}), and also
considered the effect of different bar pattern speeds on
the degree of chaoticity as quantified by the Lyapunov exponents (see
Appendix~\ref{sec:appendix:patternspeedangle}). Since the stars in our sample phase-mix quickly, the bar orientation should not have an effect (we verified that this is indeed the~case). 

We find that a less massive boxy bar of a similar extent as the
fiducial bar potential does not yield significantly different 
results. A shorter bar, extending only to a few kpc, causes little
chaoticity for stars on prograde orbits, and only affects radial
orbits with $|L_z| \lesssim 500 $ kpc km s$^{-1}$, but causes a larger
and higher degree of chaoticity (see Fig.~\ref{fig:Lyap_IoM_otherbar}).

Regarding the dependence on the pattern speed, we find that our results are robust under reasonable
variations, with at most 2\% differences in the number of stars having
$\lambda = 0$ (with the number of regular orbits being slightly
larger for lower pattern speeds), which is consistent with literature
\citep{Manos2011}. 
A larger bar pattern speed causes the
distribution's peak to shift towards higher Lyapunov exponents and
thus shorter Lyapunov timescales (see Fig.~\ref{fig:vary_patternspeed_angle}). This is to be expected, as the bar's
time-dependent perturbation literally increases its frequency.

Finally, throughout this work we have assumed a spherical dark matter
halo, but there is an increasing body of evidence suggesting that it is triaxial
\citep{LawMajewski2010, VasilievTango2021, Woudenberg2024}. We have explored 
the effect of a triaxial halo by
taking the fiducial bar potential and replacing the spherical DM halo
by the best-fit halo found by \cite{Woudenberg2024}, and found no significant changes to our results.

\subsection{The blue and red halo}

It has recently been established \citep{GaiaDR2_CMD_2018} that the colour-(absolute) magnitude diagram of halo stars near the Sun shows two sequences: a red sequence, corresponding to stars on \enquote{hot} thick disk like orbits, and a blue sequence, corresponding to accreted stars \citep{Helmi2018, Gallart2019}. The stars in the two sequences follow different phase-space distributions, and it is interesting to see whether they are affected differently by the bar because of this. Stars in either sequence can be selected by employing a single isochrone that splits the two sequences. In this work, we followed \cite{Dodd2024_SFH} and used a BaSTI-IAC alpha-enhanced isochrone with an age of 11.6 Gyr and $\rm [M/H]=-0.509$, which corresponds to $\rm [Fe/H]\sim-0.8$~dex. In this way, we find 14206 stars in the blue sequence and 13679 in the red sequence.

Using the definitions introduced earlier to identify chaotic orbits, namely $\lambda > 0$ and \fdr > -1.9, we find that in the fiducial rotating bar potential, 57\% of the blue halo stars are on chaotic orbits, versus 63\% for those on the red halo sequence. In the static bar potential, these percentages are 53\% versus 59\% for the blue and red halo sequences, respectively. 

Fig. \ref{fig:bar_resonances_hist} quantifies the percentage of stars in the red and blue halo sequences close to different bar resonances. Although the total percentage for both is similar, adding up to roughly 13\%, individual resonances are populated differently due to the difference in the red and blue halo stars' phase-space distributions. While the red halo, for example, populates the $l/m =0$ and $l/m = 1/4$ resonance more by factors of 2.2 and 2.8, respectively, the blue halo populates the $l/m = 1/2$ and  $l/m = 3/2$ resonance more by a factor 1.4 and 3.2, respectively. The main reason for the difference in degree of chaoticity however, seems to be due to the red halo having a greater percentage of stars on highly bound $(E < -150 000$ km$^2$ s$^{-2}$) orbits, where largest number and most chaotic orbits are found, and also where the corotation resonance is located for our potential.

\subsection{Limitations and implications}

Throughout this work, we have made some simplifying assumptions. Most importantly, we have assumed that the bar has fixed parameters. However, recent work suggest the bar has slowed down in the past Gyrs \citep{Chiba2021}, which would result in changes in the resonance structure with time. However, the rate of change inferred is typically larger than the orbital timescales for the stars in our sample, and could thus be considered to be an adiabatic perturbation. More importantly, we have shown in Sect.~\ref{sec:results:degreechaos} that most of the chaoticity is driven by the triaxial shape of the bar, rather than by its rotation, so most of the results presented here should hold. 

We have also neglected the perturbations from the infall of the LMC \citep{GaravitoCamargo2019, Vasiliev2023_LMC}, but this will likely mostly affect stars with large apocenters, on which the bar has no significant effect unless they are extremely radial. Lastly, we have neglected the perturbations from the spiral arms, but as these might be transient and limited to very low $z$, they are dynamically not of great importance for our sample of halo~stars. 

In this work, we have considered a relatively small local volume ($d < 1$ kpc) to ensure completeness. This also implies that we have obtained a heliocentric perspective, as the effect of the bar will differ depending on location in the Galaxy. We observe that a small fraction of stars, mainly those on highly bound and radial orbits, can shrink their apocenter to values below 7~kpc within a few Gyr due to the bar. Hence, also in this sense, the observed distribution in $(E, L_z, L_{\bot})$ space should be taken as a snapshot taken at the present-time. 

The high degrees of chaoticity found in this work imply that some
structures, depending on their present-day location in ($E$, $L_z$,
$L_\perp$), may have phase-mixed faster than previously thought
\citep{HelmiWhite1999}. Spatially coherent stellar streams may also
have been affected by the bar. In fact, chaoticity has been put
forward as a hypothesis to explain the short length of the Ophiuchus
stream \cite{PriceWhelan2016bar, Yang2025}, and it would be
interesting to explore the presence of such signatures for other
streams whose orbit might be chaotic. Next, we find that stars on chaotic
orbits move along stripes with positive slope in $(E, L_z)$ space when
close to a bar resonance. Hence, a clump of stars in
$(E, L_z, L_{\bot})$ space, if sufficiently large, may spread out along such a stripe over time.

We also observe the presence of a large number of bar resonances in
the stellar halo. Obviously, the exact locations and number of stars
close to bar resonances in $(E, L_z, L_{\bot})$ space depends strongly
on the exact configuration of the potential. Although the percentage
of stars on bar resonances is relatively small, of the order of a few
percent, these can cause overdensities, whose interpretation should be conducted with
care. Only if kinematic features are detected that seem to be introduced by the bar, one can use these to constrain a bar model, as recently attempted by \cite{Dillamore2025}. 

\section{Conclusion}
\label{sec:conclusion}

In this work, we have explored the effect of a rotating stellar bar on the orbits of halo stars located within 1 kpc from the Sun, and more specifically the effect on their distribution in $(E, L_z, L_{\bot})$ space. We have characterised their orbits using orbital frequencies and two chaoticity indicators, the Lyapunov exponent $\lambda$ and the frequency diffusion rate \fdr.  We selected stars close to bar resonances and mapped their location in  $(E, L_z, L_{\bot})$ space. 

Using as criteria for the identification of truly chaotic orbits that $\lambda > 0$ and \fdr > -1.9, we find that, while in an axisymmetric potential only $\sim 10\%$ of the orbits is strongly chaotic, in a static bar potential this increases to 56\%, and in a rotating bar potential this increases to 60\%. This fraction is not strongly dependent on the bar pattern speed, bar orientation, or other of its properties. Stars on the red halo sequence are slightly more often on chaotic orbits than those in the blue halo sequence for a range of rotating bar potentials. This is driven by their different phase-space distribution. We find that a large percentage of stars on bound, radial and prograde orbits are chaotic, interspersed with regions of more stable orbits corresponding to the locations of bar resonances. We find that the main driver of chaoticity is the bar's triaxial shape, and that the importance of bar resonances (and their overlap) as a driver of chaotic evolution appears to be secondary. 

We quantify the impact of the chaoticity over a Hubble time by computing the Lyapunov timescale $t_{\lambda}$ (including a chaos onset time). We find that most stars with high binding energy and $ |L_z| \lesssim 1000$ kpc km s$^{-1}$ are on chaotic orbits that have Lyapunov timescales that are much shorter than a Hubble time. Stars may move over time along a line in $(E, L_z, L_{\bot})$ space defined by bar resonances. Orbits can change from prograde to retrograde, from disk-like to reaching high above the midplane, and their apocenters can grow or shrink. On the other hand, we find that stars with lower binding energy ($\gtrsim - 80 000$ km$^2$ s$^{-2}$), or which reach high above the Galactic plane ($L_{\bot}$ $\gtrsim 1500$ kpc km s$^{-1}$), or which are highly retrograde ($L_z \lesssim -1000$ kpc km s$^{-1}$), have Lyapunov timescales longer than a Hubble time and do not change their  $(E, L_z, L_{\bot})$ significantly. 

As a consequence of the above, the substructures identified as potentially
being associated to merger debris and located in these specific regions of
$(E, L_z, L_{\bot})$ become more spread out. We have explored this for
the substructures identified by \cite{Dodd2023}, and we found important
changes that may lead to overlap for substructures that are on low
inclination prograde orbits, as well as for those that are strongly
bound and on radial orbits.  This leads us to propose the use of
$(E_J, L_z, L_{\bot})$ space in future clustering analysis, as $E_J$
is a conserved quantity in a rotating bar potential (and will change
only adiabatically in case the bar slows down) and substructures
separate well in this space. We find that substructures on less bound
orbits remain fairly coherent even in the presence of the bar.

These results likely also have implications for Galactic globular clusters and their possible associations to different merger events \citep[see for example][]{Massari2019}. Interestingly, in recent work, \cite{GC2025} find comparable fractions of globular clusters on chaotic and on regular orbits as found for our local sample of halo stars, even though they probe rather different regions of the Galaxy.


\begin{acknowledgements}
This research has been supported by a Spinoza Grant from the Dutch Research Council (NWO). HCW thanks Eugene Vasiliev for sharing his expertise and his help with a range of questions on AGAMA, and Emma Dodd for helpful discussions.
This work has made use of data from the European Space Agency (ESA) mission
{\it Gaia} (\url{https://www.cosmos.esa.int/gaia}), processed by the {\it Gaia}
Data Processing and Analysis Consortium (DPAC,
\url{https://www.cosmos.esa.int/web/gaia/dpac/consortium}). Funding for the DPAC
has been provided by national institutions, in particular the institutions
participating in the {\it Gaia} Multilateral Agreement.

Throughout this work, we have made use of the following packages: \texttt{astropy} \citep{Astropy},
          \texttt{vaex} \citep{vaex2018},
          \texttt{SciPy} \citep{2020SciPy-NMeth},
          \texttt{matplotlib} \citep{matplotlib},
          \texttt{NumPy} \citep{Numpy},
          \texttt{AGAMA} \citep{AGAMA}, 
          \texttt{SuperFreq} \citep{SuperFreq},
          \texttt{tqdm} \citep{tqdm},
          and Jupyter Notebooks \citep{JupyterNotebook}.
\end{acknowledgements}

\bibliography{bibliography-ah}
\bibliographystyle{aa} 

\begin{appendix}
\label{sec:appendix}

\section{Different bar models}
\label{sec:appendix:barmodels}

The length, mass, and mass distribution of the bar are uncertain. The \cite{Sormani2022} bar model is relatively heavy and long, reaching a radius of about 5~kpc. However, recent work suggests that the length of the bar might be overestimated, as stars part of an attached spiral arm are included, which would bring the length of the bar down to about 3.5~kpc \citep{Lucey2023bar, Vislosky2024}. As the bar temporarily overlaps with the inner spiral arms, it can appear up to twice its true size \citep{Hilmi2020}. To investigate how having a less massive and shorter bar with a different mass distribution would change our results, we took the fiducial potential from \cite{Woudenberg2024} with a spherical NFW halo (for consistency with the fiducial potential) and replaced its bulge by a bar of the same mass ($8.9 \cdot 10^9 M_{\odot}$). This bar is a boxy bar of the form 

\begin{equation}
    r = \left( \left(\frac{x}{a}\right)^k + \left(\frac{y}{b}\right)^k + \left(\frac{z}{c}\right)^k \right)^{1/k}
    \label{eq:rdensityprofilebar}
\end{equation}

\begin{equation}
    \rho(r) = \rho_0 e^{-r^{1/n}}
    \label{eq:densityprofilebar}
\end{equation}

\noindent with $k =4$. We kept the pattern speed and bar angle the same as in the fiducial potential, and investigated two different bar parametrizations. Firstly, we investigated a boxy bar that has a similar length as the \cite{Sormani2022} bar, with parameters $n = 1.1$, $a = \frac{7}{10}$, $b = \frac{7}{40}$, $c = \frac{7}{80}$ and $\rho_0 = 1.4 \cdot 10^{10} M_{\odot}$ kpc$^{-3}$. Secondly, we investigated a short boxy bar with a much smaller extent than the \cite{Sormani2022} bar, about 2.5~kpc, with parameters $n = 0.7$, $a = \frac{7}{10}$, $b = c = \frac{7}{40}$, and $\rho_0 = 2.9 \cdot 10^{10} M_{\odot}$ kpc$^{-3}$. We checked that the rotation curves of these two models agree with the rotation curve data by \cite{Eilers2019} and \cite{Ou2023}. 

\begin{figure*}
\centering
\begin{subfigure}{0.97\textwidth}
    \centering
    \includegraphics[width=.95\linewidth]{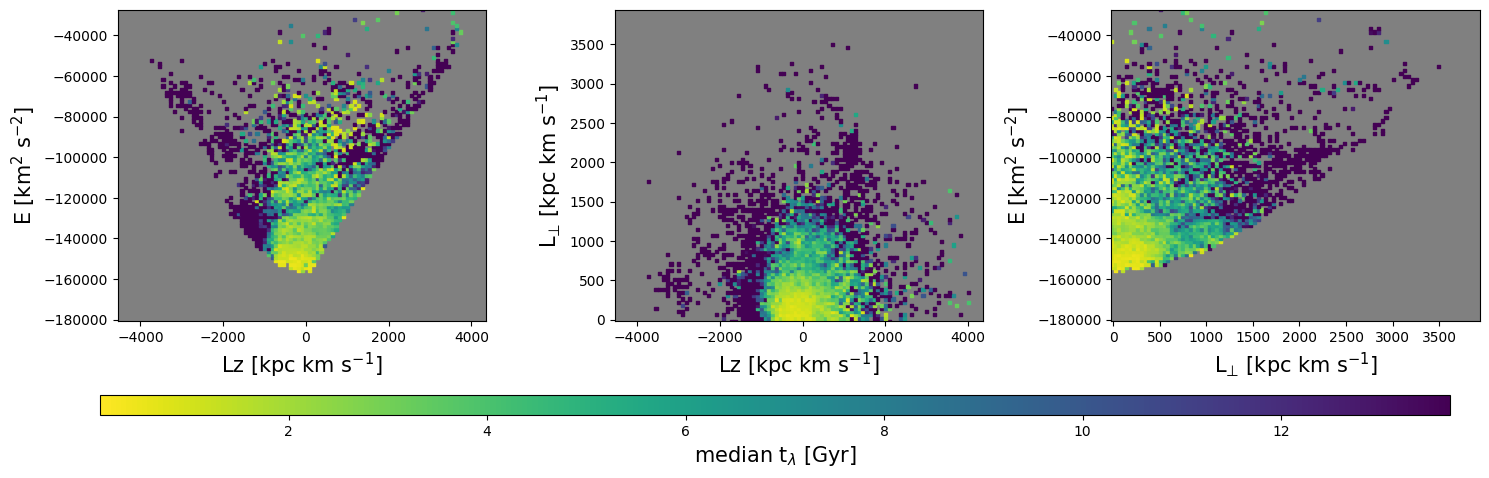}  
\end{subfigure}
\begin{subfigure}{0.97\textwidth}
    \centering
    \includegraphics[width=0.95\linewidth]{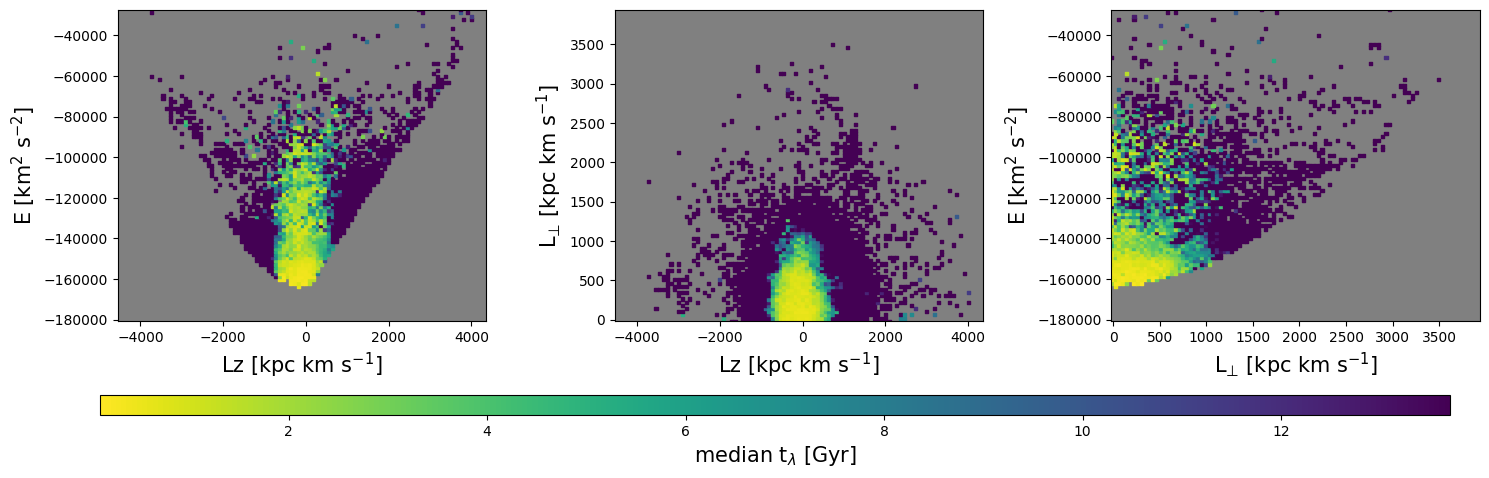}
\end{subfigure}
\caption{Local halo stars in $(E, L_z, L_{\bot})$  space at the present day, binned into 100 bins in both $x$ and $y$ direction, showing the median Lyapunov time in the boxy bar (top row) and short boxy bar model (bottom row).}
\label{fig:Lyap_IoM_otherbar}
\end{figure*}

The distribution of Lyapunov timescales in IoM space is shown in Fig. \ref{fig:Lyap_IoM_otherbar}. The distribution and amount of stars close to and on bar resonances for the boxy bar resemble the results we found for the fiducial potential, though of course their exact location in $(E, L_z, L_{\bot})$  space is shifted (for example, the corotation resonance shifted towards $L_z \sim 0$ kpc km s$^{-1}$ and more bound orbits). The percentage of stars with Lyapunov exponent > 0 and \fdr > -1.9 is 61\% (with 63\% in the red halo and 60\% in the blue halo). 66\% of stars with $|L_z| < 500$~kpc km s$^{-1}$ are chaotic. Instead, for the short boxy bar, the percentage of stars close to bar resonances is halved, and only stars in the range $-500 \lesssim L_z  \lesssim \sim 500$ kpc km s$^{-1}$ are for a large percentage on chaotic orbits (which are generally more chaotic, meaning having higher Lyapunov exponents and \fdr, than for the boxy bar or fiducial potential). The most bound stars in the short boxy bar model are for a larger percentage chaotic. We found that the \fdr determined using the frequencies in Poincaré symplectic coordinates in the rotating frame was more reliable than in the Cartesian rotating frame. Using that, the percentage of stars with Lyapunov exponent > 0 and \fdr > -1.9 is 60\% (with 63\% in the red halo and 57\% in the blue halo). 81\% of stars with $|L_z| < 500$~kpc km s$^{-1}$ are chaotic. Although the location of chaotic orbits is slightly different, for both models, the percentage of stars on chaotic orbits agrees with what has been found for the fiducial potential (see Sect. \ref{sec:results:degreechaos}). 

\section{Effect of different pattern speeds and bar orientation on the Lyapunov exponent}
\label{sec:appendix:patternspeedangle}

\begin{figure*}[t!]
\centering
    \includegraphics[width=\hsize]{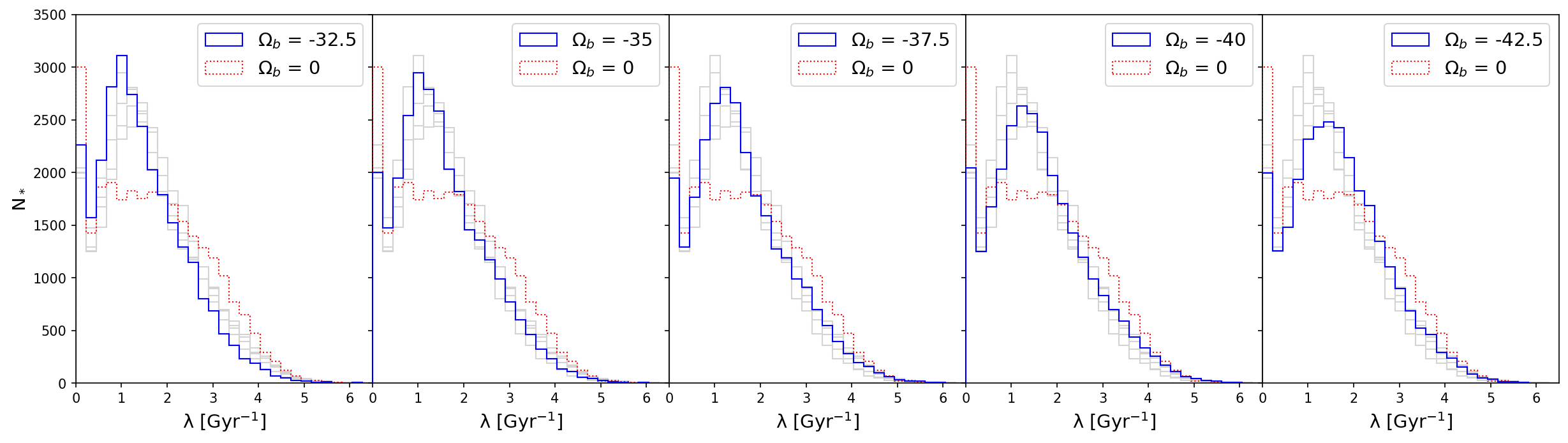}
\caption{Lyapunov exponent distributions for the local halo sample in the fiducial potential with various bar pattern speeds, $\Omega_b =$[~-~32.5, -35, -37.5, -40, -42.5] $\si{\: km \: s^{-1} kpc^{-1}}$, at a fixed bar angle, -25 degrees. The grey lines show all cases, while the blue lines show one pattern speed as indicated in the legend. For comparison, the red dotted line shows the distribution for a static bar ($\Omega_b = 0$).
\label{fig:vary_patternspeed_angle}}
\end{figure*}

To see what influence a different pattern speed has on the presence of chaoticity, we took the fiducial potential and integrated orbits of the 1~kpc volume halo sample as done before. We computed the Lyapunov exponent for various bar pattern speeds, $\Omega_b =$ [~-32.5, -35, -37.5, -40, -42.5] $\si{kpc \: km \: s^{-1}}$ at a fixed bar angle, -25 degrees. The bar angle does not influence the Lyapunov distributions (we ensured this is indeed the case), as the sample we're considering is phase-mixed. We also compared this to the distribution of Lyapunov exponents when the bar is static ($\Omega_{B}$~=~0). The results are shown in Fig. \ref{fig:vary_patternspeed_angle}, which shows for a higher bar pattern speed higher values of the Lyapunov exponent are reached. This is consistent with the findings of \cite{PriceWhelan2016bar}, see their Fig. 5. When the bar is static, there is a larger percentage of stars on regular orbits (meaning with $\lambda$ = 0).

\end{appendix}

\end{document}